\documentclass[onecolumn,preprintnumbers,amsmath,prc,amssymb]{revtex4}

\usepackage{graphicx}
\usepackage{dcolumn}
\usepackage{bm}
\usepackage{amsmath,amssymb}
\usepackage{lipsum}
\usepackage{slashed}
\usepackage{enumitem}
\usepackage{txfonts}
\usepackage{esint}
\usepackage{xcolor}
\usepackage{mathtools}
\usepackage{transparent}
\usepackage{upgreek}

\begin{document}
\title{Cancellation of the sigma mode in the thermal pion gas by quark Pauli blocking }

\author{David Blaschke}
\email{david.blaschke@uwr.edu.pl}
\address{Institute of Theoretical Physics, University of Wroclaw, Max Born Pl. 9, 50-204 Wroclaw, Poland}

\author{Alexandra Friesen}
\email{avfriesen@theor.jinr.ru}
\address{Joint Institute for Nuclear Research, 141980, Dubna, Russia}
\address{Dubna State University, 141982, Dubna, Russia}

\author{Yuriy Kalinovsky}
\email{kalinov@jinr.ru}
\address{Joint Institute for Nuclear Research, 141980, Dubna, Russia}
\address{Dubna State University, 141982, Dubna, Russia}


\begin{abstract}
We calculate the pressure of the interacting pion gas using the Beth-Uhlenbeck approach to the relativistic virial expansion with Breit-Wigner phase shifts for the $\sigma$- and $\varrho$- meson resonances. The repulsive phase shift $\delta^2_0$ is taken from quark interchange model of Barnes and Swanson [Phys. Rev. D 46 (1992) 131] in very good agreement with experimental data. In this work we show that the cancellation of the attractive (I = 0) and repulsive (I = 2) isospin channel contributions to the scalar $\pi-\pi$ interaction in the low-energy region that is known for the vacuum phase shifts, takes place also at finite temperature. This happens despite the strong medium dependence of these phase shifts that enters our model by the temperature dependence of the $\sigma$- meson and constituent quark masses because for these masses the relation $M_\sigma(T) \approx 2 m (T)$ holds and the scattering length approximation is valid as long as the strong decay channel $\sigma \to \pi \pi$ is open. 
Exploiting the Nambu--Jona-Lasinio model for describing the dynamical breaking of chiral symmetry in the vacuum and its restoration at finite temperature, we justify with our approach that the $\sigma$-meson should be absent from the hadron resonance gas description at low temperatures because the above cancellation holds. However, since this cancellation breaks down in the vicinity of the hadronization transition, where due to chiral symmetry restoration the decay channel $\sigma \to \pi \pi$ closes and the $\sigma$- meson becomes a good resonance, the latter should be included into the statistical model description of chemical freeze-out in heavy-ion collisions.
\end{abstract}


\maketitle

\date{\today}

\section{Introduction}

The thermodynamics of the hadron resonance gas (HRG) is of crucial importance for the interpretation of the results of the ab-initio evaluation of the QCD partition function by  simulations of the lattice gauge theory as well as for the explanation of yields of hadrons produced in ultrarelativistic heavy-ion collisions at the chemical freeze-out.
This HRG model makes the simplifying assumption that instead of accounting for the interactions among hadrons one may just evaluate the statistical sum over all known resonances in the spectrum of hadrons as eigenstates of the QCD Hamiltonian. While such a procedure would account for the attractive interaction leading to the formation of resonances in the spectrum, one may additionally account for repulsion by invoking an excluded volume of the hadrons treated, e.g., as hard spheres \cite{Venugopalan:1992hy}.

A simple model system for the study of the interplay between attraction and repulsion in hot hadronic matter is the interacting pion gas at finite temperature. For its description one can use the Beth-Uhlenbeck approach with the well measured phase shifts of the pion-pion interactions in free space that allow to evaluate the second virial coefficient of the partition function
\cite{Welke:1990za}.
It has soon be realized \cite{Venugopalan:1992hy} that the contributions from phase shifts in the attractive isospin-zero $\sigma$-meson channel ($\delta_0^0$) and in the repulsive isospin-2 channel ($\delta_0^2$) largely compensate each other in the partition function. 
This cancellation has also been obtained within the Nambu-Jona-Lasinio (NJL) model description of pion-pion scattering lengths at finite temperature \cite{Quack:1994vc} where it could be traced to the manifest chiral symmetry in this field-theoretic quark model of the interacting pion gas.
This model reproduces the Weinberg relations for the scattering lengths which are the basis for the cancellation.
On the quark level of description, the repulsion in the $\pi\pi$ scattering is due to the quark exchange interaction between pions, represented on the quark one-loop level by the so-called box diagrams. 
It can be shown that this quark exchange interaction, also denoted as quark Pauli blocking, leads to a repulsive phase shift well in accordance with 
the experimental data \cite{Barnes:1991em}.

The fact of the cancellation of the scalar isoscalar $\sigma$-meson (f$_0$(500)) against the repulsion in the scalar isotensor channel has recently been rediscovered in \cite{Broniowski:2015oha} when it was discussed whether the $\sigma$ meson should be included to the HRG thermodynamics.
The authors concluded that one should {\it not} incorporate f$_0$(500) in standard HRG models for studies of isospin averaged quantities \cite{Broniowski:2015oha,Giacosa:2016rjk}. 
They also suggest that a similar cancellation should take place in the corresponding channels for kaon interactions with the $\kappa$ as the chiral partner state of the kaon, see also \cite{Friman:2015zua} for a formulation within the S-matrix approach \cite{Dashen:1969ep}.  

On the other hand, in \cite{Andronic:2008gu} it was pointed out that the inclusion of the $\sigma$-meson to the HRG gives a significant improvement of the statistical model description of the ''horn'' structure in the beam energy dependence of the kaon to pion ratio $K^+/\pi^+$.   
The question arises whether both statements can be true simultaneously. 
Namely, that on the one hand in the Beth-Uhlenbeck (or S-matrix) formulation of the thermodynamics of the interacting HRG a cancellation of the attractive $\delta_0^0$ channel ($\sigma$-meson) against the repulsive $\delta_0^2$ channel (Pauli blocking) occurs, and that on the other hand a sigma meson has to be considered as an important degree of freedom in the statistical model of particle production where a sudden (chemical) freeze-out of particle species  
occurs in the vicinity of the QCD chiral restoration/quark deconfinement temperature.

We want to answer this question affirmative in the following way.
When formulating the hadron resonance gas in the S-matrix formalism one has to take into account the medium effects on the scattering phase shifts, i.e. to apply the generalized Beth-Uhlenbeck approach.
This should then in particular take into account the effects of the chiral symmetry restoration, namely that the sigma meson becomes degenerate with the pion (parity doubling) above the chiral restoration transition that entails dropping quark masses. 
Before the sigma meson mass becomes degenerate with that of the pion it has to cross the two-pion mass threshold where the strong two-pion decay channel of the sigma meson closes and it becomes a sharp resonance \cite{Volkov:1997dx}. 
For a resonance at the threshold the scattering length approximation breaks down, while a Breit-Wigner ansatz for the phase shifts will be appropriate.
The dropping quark masses entail a Mott effect for both, the pion and the sigma. This character change of the mesons from bound states to resonances in the continuum leads to the ceasing of the quark Pauli blocking effect and thus the resulting phase shift should turn to zero. 

So, the cancellation of contributions from the S-wave resonances in the attractive (I = 0) and the repulsive (I = 2) channels to the thermodynamics of the pion gas discussed on the basis of the Beth-Uhlenbeck approach to the second virial coefficient which uses the known free-space phase shifts $\delta_l^I(\sqrt{s_{\pi\pi}})$ in the dominant channels $(l,I)=(0,0), (1,1), (0,2)$ can be applicable also at finite temperature. This finite-temperature cancellation can be seen in the framework of the NJL model \cite{Quack:1994vc}, where the pion-pion scattering lengths were obtained at finite temperature. 

The goal of this work is to show how the $\sigma$-meson cancellation works at finite temperature when one uses
the NJL model results for the temperature dependence of the $\sigma$-meson mass and width parameter in the Breit-Wigner ansatz for $\delta_0^0$ phase shift on the one hand and the temperature dependent quark mass in the non-relativistic quark exchange model for $\delta_0^2$ channel on the other. 

The paper is organised as follows. 
We start from the Breit-Wigner phase shifts for the $\delta_1^1$
($\rho$-meson) and $\delta_0^0$ ($\sigma$-meson) channels \cite{Welke:1990za} and use the result of the nonrelativistic potential model calculation of the repulsive $\delta_0^2$ channel from the quark-exchange Born diagrams  \cite{Barnes:1991em} that takes into account the quark substructure of the pion. 
In Sec.~\ref{NJL_section} the temperature dependence of the mass spectra and phase shifts is calculated in the framework of the 
NJL model. 
The results are presented and discussed in Sec.~\ref{result}.

\section{The interacting pion gas}

\subsection{Beth-Uhlenbeck equation and cancellation of the $\sigma$ meson}
The virial expansion of the grand canonical partition function of the system with the known S-matrix for two-particle scattering  can be written as a virial expansion \cite{Dashen:1969ep,Venugopalan:1992hy}
\begin{equation}
{\rm ln}Z = {\rm ln} Z_0 + \sum_{i_1,i_2} z_1^{i_1} z_2^{i_2} b(i_1, i_2),
\label{Zfunc}
\end{equation}
where  $z_j = \exp(\beta\mu_j)$ for $j = 1,2$ and $b(i_1, i_2)$ is
the second virial coefficient defined by the S-matrix with labels $i_1, i_2$ referring to a channel of the S-matrix with $i_1+i_2$  particles in initial state.
\begin{equation}
b(i_1, i_2) = \frac{V}{4 \pi i}\int\frac{d^3 P}{(2\pi)^3}\int d\epsilon\exp(-\beta\sqrt{P^2+\epsilon^2}){\rm Tr}_{i_1,i_2}\left[AS^{-1}\frac{\overleftrightarrow \partial }{\partial\epsilon}S\right].
\label{EqVirKoef}
\end{equation}
Here $\beta= T^{-1}$ is the inverse temperature, V is the volume, P is the centre-of-mass (total) momentum and $\epsilon$ the energy of the two-particle system. The symbol $A$ denotes the symmetrization/antisymmetrization operator for a system of bosons/fermions.  
The trace is taken over all combinations of particle number.  
For a one-component system under the assumption that hadrons interact mainly via elastic collisions, the second virial coefficients of Eq.~(\ref{EqVirKoef}) can be simplified  by chosing the representation of the S-matrix in terms of the two-particle phase shifts. 
The lowest virial coefficient $b_2$ corresponds to the case $i_1=i_2=1$. Then Eq.~(\ref{EqVirKoef}) can be written as
 \begin{equation}
 b_2=\frac{1}{2\pi^3\beta}\int_M^\infty d\epsilon\epsilon^2 K_2(\beta\epsilon)\sum_{l,I}{'}g_{l,I}\frac{\partial\delta_l^I(\epsilon)}{\partial\epsilon},
 \label{eq_b2}
 \end{equation}
with the modified Bessel function $K_2$ 
and the degeneracy factor $g_l^I=(2l+1)(2I+1)$.
$M$ is the invariant mass of the interacting pair at threshold. 
For given $l$ the sum over $I$ is restricted to values consistent with statistics. 
In the limit that only the second virial coefficient is considered, the interaction pressure (as well as all the other thermodynamic variables) can be obtained from  Eq. (\ref{Zfunc}) in the form of a Beth-Uhlenbeck equation,
\begin{equation}
P_{int} = T z^2 b_2 = \frac{z^2}{2\pi^3\beta^2}\int_M^\infty d\epsilon\epsilon^2 K_2(\beta\epsilon)\sum_{l,I}{'}g_{l,I}\frac{\partial\delta_l^I(\epsilon)}{\partial\epsilon}.
\label{Pint}
\end{equation}
For the one component pion gas, the total center of mass energy is chosen as  $\epsilon=2 (q^2+ M_\pi^2)^{1/2}$ and the threshold mass is $M = 2 M_\pi$ in Eqs.~(\ref{eq_b2}) and (\ref{Pint}). 
For the case $\delta_l^I\rightarrow 0$ at low energies $\epsilon\rightarrow M$  the equation for the second virial coefficient $b_2$  can be simplified using integration by parts to
\begin{eqnarray}
 \label{b2}
 b_2 &=&\frac{1}{2\pi^3}\int_{2M_\pi}^\infty d\epsilon\epsilon^2 K_1(\beta\epsilon)\sum_{l,I}{'}g_{l,I}\delta_l^I(\epsilon).
\end{eqnarray}
The corresponding pressure contribution from the two-particle interactions is 
\begin{eqnarray}
 P_{int}& =&  P^{00}+P^{02}+P^{11}~,
 \label{eq:Pint}
\end{eqnarray}
where the partial pressure contributions at vanishing chemical potential ($z=1$) are defined by the phase shift,                                                                                                       
\begin{eqnarray}
 P^{lI}& =&  \frac{g_{l,I}}{2\pi^3\beta}\int_{2 M_\pi}^\infty d\epsilon\epsilon^2 K_1(\beta\epsilon)\delta_l^I(\epsilon).
 \label{press_b1}
\end{eqnarray}

\begin{figure}[ht]	
\centerline{
\includegraphics[width=5.05cm]{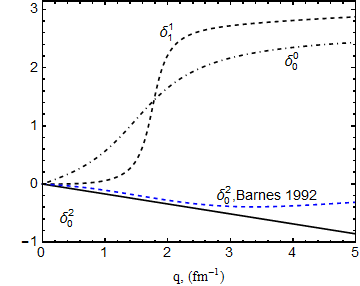}
\includegraphics[width=6cm]{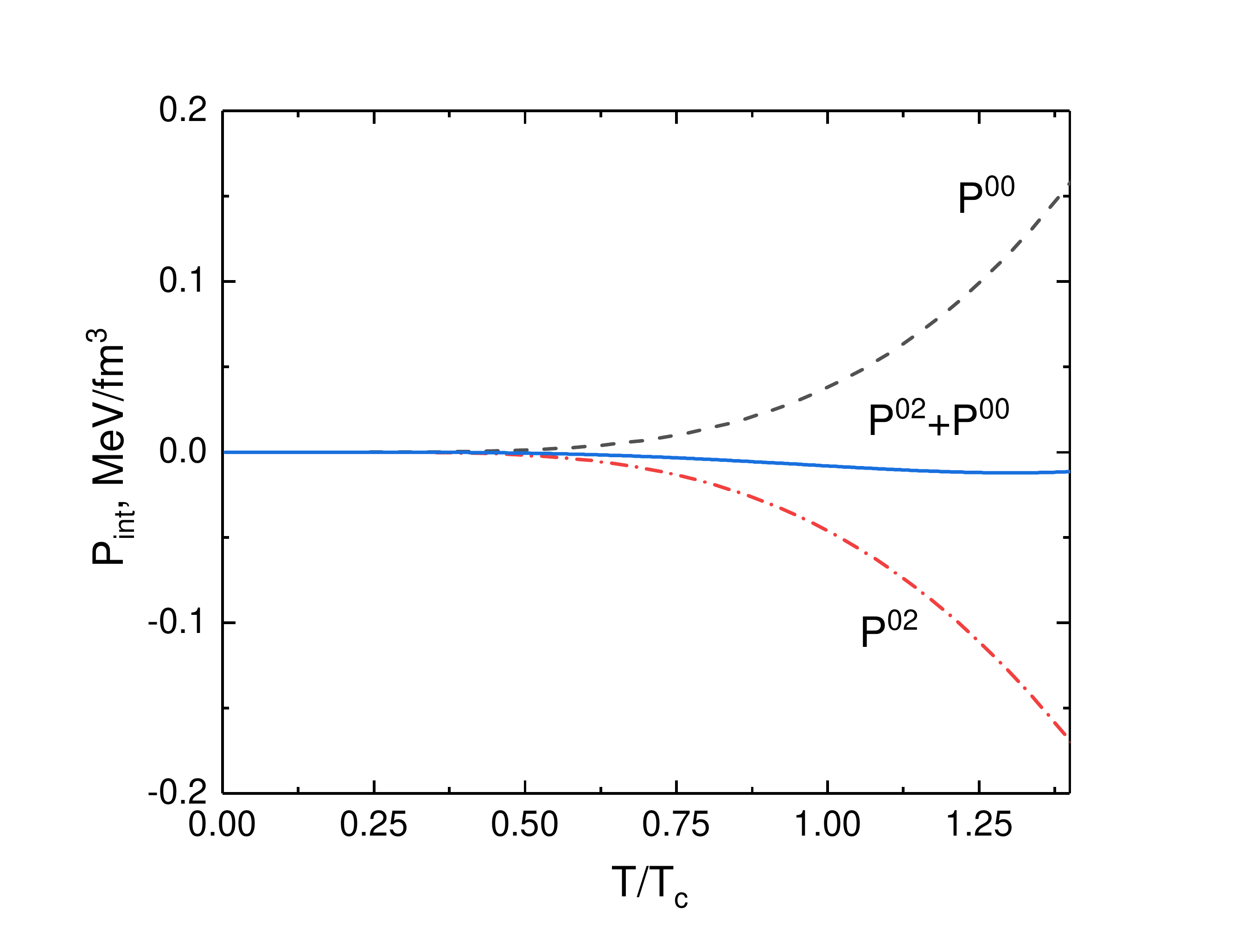}
}
\caption{Left panel: $\pi\pi$-scattering phase shifts as functions of the total center-of-mass-momentum. The blue dashed line corresponds to the phase shift $\delta_0^2$ calculated using Eq.~(\ref{d02Barnes}). 
Right panel: Interaction contribution to the pressure according to the Beth-Uhlenbeck formula with the medium-independent phase shifts of the left panel (without the contribution from the $\rho$-meson channel).  The almost perfect cancellation of 
the $\sigma$-meson (green dashed line) by the quark Pauli blocking (red dashed-dotted line) is demonstrated by the blue solid line.
\label{phaseshifts_q}}
\end{figure}

The second virial coefficient $b_2$ and therefore the thermodynamic functions are . 
The phase shift in free space (vacuum) can be obtained from experiments and be compared with theoretical models for it. 
The latter could then also be used to predict its in-medium modifications.
The low-energy $\pi\pi$ interaction includes the $\delta_0^0,\delta_1^1, \delta_0^2$ phase shifts. 
The repulsive isospin-2, S-wave phase shift 
can be well described in the scattering length approximation by $\delta_0^2 = -0.12q/M_\pi$.  
The phase shift  $\delta_1^1$ contains the $\rho$-meson resonance and $\delta_0^0$ contains the $\sigma$-meson resonance. 
These resonant phase shifts can be chosen in the simple Breit-Wigner form \cite{Welke:1990za},
\begin{eqnarray}
\delta_0^0(\epsilon)&=& \frac{\pi}{2}+\arctan\left(\frac{\epsilon-M_\sigma}{\Gamma_\sigma/2}\right),\label{phaseshift00}\\
\delta_1^1(\epsilon)&=& \frac{\pi}{2}+\arctan\left(\frac{\epsilon-M_\rho}{\Gamma_\rho/2}\right).
\label{phaseshift11}
\end{eqnarray}
For the case $M_\pi=0.138$ GeV, $M_\sigma=5.8~M_\pi$, $M_\rho=5.53~M_\pi$ and
\begin{eqnarray}
\Gamma_\sigma &=& \beta_\sigma q,\label{gsi_prakash} \\ 
\Gamma_\rho &=& \beta_\rho q \left(\frac{q/M_\pi}{1+(q/M_\rho)^2}\right)^2, \label{gro_prakash}
\end{eqnarray}\
with $\beta_\sigma=2.06$ and $\beta_\rho = 0.095$. 
In this approximation, the well-known picture for the phase shifts $\delta_0^0,\delta_1^1, \delta_0^2$  is obtained, see Fig.~\ref{phaseshifts_q}.

In the work \cite{Barnes:1991em} the phase shift for $I = 2$ low-energy $\pi\pi$ scattering was obtained within a diagrammatic approach to the quark exchange interaction between $q\bar{q}$ mesons in the framework of a nonrelativistic potential model for their quark substructure, see also \cite{Blaschke:1992qa}. 
The phase shift, obtained by the so-called quark Born diagrams has the form
\begin{equation}  
\sin\delta_0^2 = -\left\lbrace \frac{\alpha_s}{9\lambda m^2}{\frac{\sqrt{q^2+M_\pi^2}}{q}}\left(1-e^{-2\lambda q^2}+\frac{16\lambda q^2}{3\sqrt{3}}e^{-4 \lambda q^2/3}\right)\right\rbrace.
\label{d02Barnes}
\end{equation}
The parameter $\alpha_s$ is fixed to the value $\alpha_s/m^2 = 4.48$ GeV$^{-2}$ so that 
$\alpha_s = 0.71$ for $m=0.40$ GeV.  
The parameter $\lambda$ is related to the Gaussian wave function of the nucleon in the usual simple-harmonic-oscillator (SHO) model with the quark-model parameter $\beta_{SHO} = 1/(2\sqrt{\lambda})=0.337$ \cite{Barnes:1991em}. 
The resulting phase shift $\delta_0^2$ is shown in Fig.~\ref{phaseshifts_q} as a blue dashed line.

Inserting the phase shifts to the Beth-Uhlenbeck formula (\ref{press_b1}), one observes an almost perfect cancellation of the sigma meson by the quark Pauli blocking, see the right panel of Fig.~\ref{phaseshifts_q}.  
This result has been obtained under the assumption that the phase shifts are independent of the temperature which is no longer valid when approaching the pseudocritical temperature $T_c$, where chiral symmetry gets restored and the pion undergoes a Mott dissociation. 
This pion dissociation effect on the Pauli blocking phase shift $\delta_0^2$ can be accounted for by multiplying it with a factor 
\begin{equation}
c(T) = \begin{cases}
    |2m(T) - M_\pi(T)|/ |2m(0) - M_\pi(0)|, & \text{if $T<T_{{\rm Mott},\pi }$}.\\
    0, & \text{otherwise}.
  \end{cases}
\end{equation}
that quantifies the reduction of the binding energy when approaching the Mott temperature. The relation to the fourth power of the pion-quark-antiquark coupling  $g_{\pi qq}(T)= |2m(T) - M_\pi(T)|^{1/4}$ is plausible because the Pauli blocking effect is encoded in the quark box diagram for the four pion vertex, see Eq. (3.4) of Ref. \cite{Hufner:1996pq}.

\subsection{Analytic justification for the cancellation}

The fact of the  $\sigma$ cancellation at low energies and the robustness of this result at finite temperatures despite the 
in-medium modification of the phase shifts can be understood also analytically in the low-momentum expansion of the phase shifts of Eqs.~(\ref{phaseshift00}) and (\ref{d02Barnes}) that introduces the scattering lengths $a^I_l$,
\begin{equation}
\frac{d \delta^i_l}{dq}\bigg|_{q=0}\approx a^I_l.
\end{equation}
The $\sigma$ meson scattering length is then
\begin{equation}
a^0_0 = \frac{\beta_\sigma}{2(M_\sigma-2M_\pi)},~~\beta_\sigma=2.06,
\end{equation}
which becomes singular in the vicinity of the chiral transition, when 
$M_\sigma \to 2M_\pi$ and the dominant $\sigma\to 2\pi$ decay channel closes so that the $\sigma$ meson becomes a good resonance at the threshold. This is the limitation for the present considerations because the scattering length approximation breaks down in this case.

The scattering length for the phase shift of the quark exchange process (quark Pauli blocking) from the Eq.(\ref{d02Barnes}) gives
\begin{equation}
a^2_0 = - \frac{2}{9}\left(1+\frac{8}{3\sqrt{3}}\right)\alpha_s\frac{M_\pi}{m^2},
\end{equation}
so that we obtain for the relevant ratio of scattering lengths
\begin{eqnarray}
\frac{a^0_0}{5a^2_0}&=& \frac{1.03}{-2M_\pi + M_\sigma}\frac{1}{5}\left(-\frac{0.565\, \alpha_s M_\pi}{m^2}\right)^{-1} = -\frac{1.03}{-2M_\pi^2 + M_\sigma M_\pi} \frac{m^2}{2.82 \alpha_s} \nonumber\\
&=& \frac{1.03}{2(M_\pi-M_\sigma/4)^2-M_\sigma^2/8}\frac{m^2}{2.82 \alpha_s}\approx - \frac{4 m^2}{1.41\, \alpha_s M_\sigma^2},
\label{relation_d}
\end{eqnarray}
where in the last step the approximation $(m_\pi-m_\sigma/4)^2\approx 0$ has been used. 

The wanted result of the cancellation by quark Pauli blocking of the $\sigma$ meson contribution to the thermodynamics of the pion gas is obtained when $\alpha_s\sim 0.7$. 
Then, because $M_\sigma \approx 2m$, the result is $a^0_0/5a^2_0 = -0.99$, which means the total cancellation of contributions from $\sigma-$channel by the contribution from the repulsive Pauli-blocking channel. 
For the parameter set given by Welke \cite{Welke:1990za}  $\delta^0_0/5\delta^2_0 = -1.18$ and for the Weinberg results $a_0 = 0.158$ and $a_2 = -0.045$ follows $a^0_0/5a^2_0 = -0.708$ \cite{Weinberg:1966kf}. 

We observe that the result of the cancellation of the $\sigma$ meson as a resonance in the interacting pion gas by quark Pauli blocking depends only on the relation $M_\sigma=2m$ between the masses of the $\sigma$ and the quark as well as the approximate  validity of $(M_\pi-M_\sigma/4)^2\approx 0$.
Therefore, one can expect that it should hold also when going beyond the standard Beth-Uhlenbeck approach and taking into account an explicit temperature dependence of the phase shifts, as long as the above relations between quark and meson masses remain intact. 
In the following, we consider the Nambu -- Jona-Lasinio model for describing the temperature dependence of quark and meson masses as well as meson decay constants, so that the limits of the cancellation effect can be estimated.

\section{Mesons at finite temperature in the Nambu -- Jona-Lasinio model}
\label{NJL_section}

In Eqs.~(\ref{b2}) and (\ref{press_b1}) the dependence on the temperature appears due to the parameter $\beta = 1/T$ only and it is assumed there that other quantities (like masses and $\delta_l^I$) are temperature independent. 
Since the constituent quark mass in (\ref{d02Barnes}) is generated by dynamical chiral symmetry breaking in the vacuum, one expects its approximate restoration at finite temperatures (and densities). 
This effect of chiral symmetry restoration ($\chi$SR) in a 
hot and dense medium can be modeled with the Nambu-Jona - Lasinio (NJL) model \cite{RevModPhys.64.649}. 

We employ here the NJL model with two flavours of quarks 
defined by the Lagrangian
\begin{equation}
  \mathcal{L}_{\rm NJL}=\bar{q}\left(i\gamma_\mu \partial^\mu - \hat{m}_0
  \right) q+ G_s  \left[\left(\bar{q}q\right)^2+\left(\bar{q}i\gamma_5
      \vec{\tau} q \right)^2\right] -G_{\rm v}\left[(\bar{q}\gamma^\mu\tau^a q)^2+(\bar{q}\gamma^\mu\gamma^5\tau^a q)^2\right],
\label{njl}
\end{equation}
with chirally symmetric four-quark interactions in the scalar, pseudo-scalar, vector and axial-vector channels.
$G_s$ and $G_{\rm v}$ are the scalar and vector coupling constants, $\bar{q}$ and $q$ - the quark spinor fields, $\hat{m}_0$ is the diagonal matrix of the current quark mass,
$\hat{m}_0 = {\rm diag}(m_u^0, m_d^0)$ with $m_u^0 = m_d^0 = m_0$, and
$\overrightarrow{\tau}$ are the SU(2) Pauli matrices in flavor space with the components $\tau^a(a = 1,2,3)$.

In the mean-field approximation the constituent quark mass is obtained by solving the gap equation at the mean field level
\begin{eqnarray}
m = m_0 + 8 G_s N_c N_f \int_{\Lambda} \dfrac{d^3p}{(2\pi)^3}
\dfrac{m}{E_p} \left[ 1 - f(E^+) - f(E^-) \right]~,
\label{gap1}
\end{eqnarray}
where the dependence on temperature and chemical potential is modeled in the Fermi-functions 
$f(E^\pm) = (1 + e^{\beta E_p^\pm})^{-1}$ 
and the quark (antiquark) energy dispersion relation 
$E_p^\pm = E_p\pm \mu$.

Mesons are considered as quark-antiquark bound states and their properties are described by the Bethe-Salpeter equation in the pole approximation
\begin{equation}
1 - 2G_s \ \Pi_{M}(k^2)|_{k^2 = M_M^2} = 0~~,~M=\pi,\sigma,
\label{mesonmass}
\end{equation}
with the polarization operator $\Pi_{M}(k^2)$ determining the meson properties being defined as
\begin{equation}
\Pi_{M} (k^2) = i \int \frac{d^4p}{(2\pi)^4} \ \mbox{Tr}\,
\left[ \Gamma_M S(p+k) \Gamma_M S(p) \right],  
\end{equation}
where the vertex factor $\Gamma_M$ depends on the meson species $M=\pi$, $\sigma$, $\rho$ and $a_1$. 
For the pseudo-scalar $\pi$ meson $\Gamma_\pi = i \gamma_5 \tau^a $ and for the scalar $\sigma$ meson $\Gamma_\sigma = {\bf 1} \tau^a $; $S(q)$ is the quark propagator and the trace is being taken over color, flavor and spinor indices. 
For mesons at rest (${\mathbf P}=0$) in the medium, these conditions correspond to the equations:
 \begin{eqnarray}
1 + 8 G_s N_c N_f \int \frac{d^3 p}{(2\pi)^3}
\frac{1}{E_p} \frac{E_p^2}{M_\pi^2-4 E_p^2} \left( 1- f(E^+) - f(E^-) \right) &=& 0, 
\label{masspi}\\
1 + 8 G_s N_c N_f \int \frac{d^3 p}{(2\pi)^3}\frac{1}{E_p} 
\frac{E_p^2-m^2}{M_\sigma^2-4 E_p^2} \left(1-f(E^+) -f(E^-)\right) &=& 0~. 
\label{masssigma}
\end{eqnarray}

Both pion-quark $g_{\pi qq}(T,\mu)$ and sigma-quark $g_{\sigma qq}(T,\mu)$
coupling strengths can be defined from $\Pi_{M}$ by the residuum of the mass pole approximation (\ref{mesonmass})
\begin{eqnarray}
g_{Mqq}^{-2}(T, \mu) = \frac{\partial\Pi_{M}(k^2)}
{\partial k^2}\vert_{{k^2 =M_M^2} }~~,~M=\pi,\sigma.
\label{couple}
\end{eqnarray}

Generally, the pole mass equation (\ref{mesonmass}) can be extended to the vector and axial-vector case and the set of equations Eqs.~(\ref{gap1}), (\ref{masspi}), (\ref{masssigma}) together with an equation for the vector meson is solved  self-consistently.  
To simplify the calculations for the $\rho$ meson, the relations suggested in the works 
\cite{Volkov:1997dd,Volkov:2005kw,Ebert:1992ag} are used. 
The mass of the $\rho$-meson and its width can be calculated as
\begin{eqnarray}
M_\rho^2 &=& \frac{g_{\rho qq}^2}{4 G_{\rm v}}, \label{rho_mass}\\ 
g_{\rho qq} &=& \sqrt{6}g_{\sigma qq},\\
\Gamma_{\rho\pi\pi} &=& \frac{g_{\rho \pi\pi}^2}{48 \pi M_{\rho}^2}\sqrt{\left(M_\rho^2- 4 M_\pi^2\right)^3}. 
\label{rho_width}
\end{eqnarray}

The decay width $\Gamma_{\sigma\pi\pi}$ within the NJL model is defined by the triangle Feynman diagram treating the sigma meson as quark-antiquark system \cite{Zhuang:2000tz,Friesen:2011ma}: 
\begin{eqnarray} \label{Gam0}
\Gamma_{\sigma\pi\pi} = \frac{3}{2} \ \frac{(2g_{\sigma qq}  g^2_{\pi qq} A_{\sigma\pi\pi}(T, \mu))^2} {16\pi \
M_{\sigma}}
\sqrt{1 - \frac{4M_{\pi}^2}{M_{\sigma}^2}}~,
\end{eqnarray}
where the factor 3/2 takes into account the isospin conservation and  $g_{\sigma qq}$ and $g_{\pi qq}$ are coupling  constants defined from Eq. (\ref{couple}). 
The amplitude of the triangle vertex $A_{\sigma\rightarrow\pi\pi}$ is
\begin{eqnarray}
A_{\sigma\pi\pi} = \int \frac{d^4q}{(2\pi)^4} \
Tr \{S(q) \ \Gamma_\pi \ S(q+P) \ \Gamma_\pi \ S(q)\}.
\end{eqnarray}
The kinematic factor $\sqrt{1-{4M_{\pi}^2}/{M_{\sigma}^2}}$ in
(\ref{Gam0})  leads to the constraint  
$M_{\sigma}>2M_{\pi}$, if this condition is broken, the decay $\sigma\rightarrow\pi\pi$ is forbidden
and the $\sigma$ meson becomes a good bound state 
of the $\pi\pi$ interaction with only a negligible electromagnetic decay width from the $\sigma \to \gamma \gamma$ process \cite{Volkov:1997dx}.

 In order to describe the mass spectra in the NJL model, a set of  parameters is required.
 Since we are interested in a phenomenology for the case that the constituent quark mass in the vacuum is $m=400$ MeV,
 we choose the corresponding parameterization from table V of Ref.~\cite{Grigorian:2006qe} with
 the cutoff parameter $\Lambda=0.5879$ GeV, the current quark mass $m_0 = 5.582$ MeV  and the scalar-pseudoscalar coupling constant $G_s \Lambda^2= 2.442$.  
 The vector meson coupling constant  $G_{\rm v}  \Lambda^2 = 2.4$ is found from fitting the $\rho$ meson mass in vacuum $M_\rho=780$ MeV. 

\begin{figure}[t]	
\centerline{
\includegraphics[width=5.5 cm]{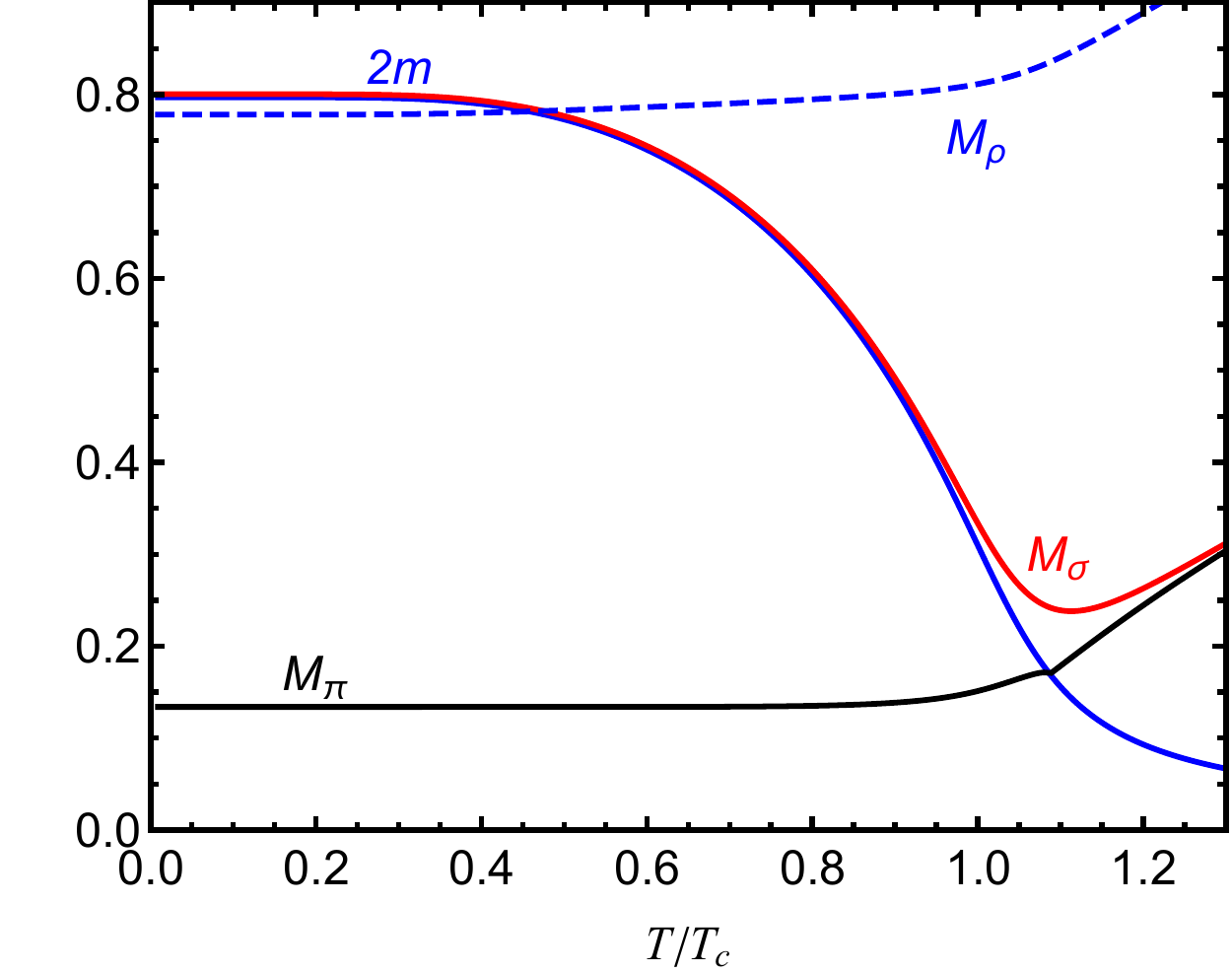}
\hspace{1cm}
\includegraphics[width=5.5 cm]{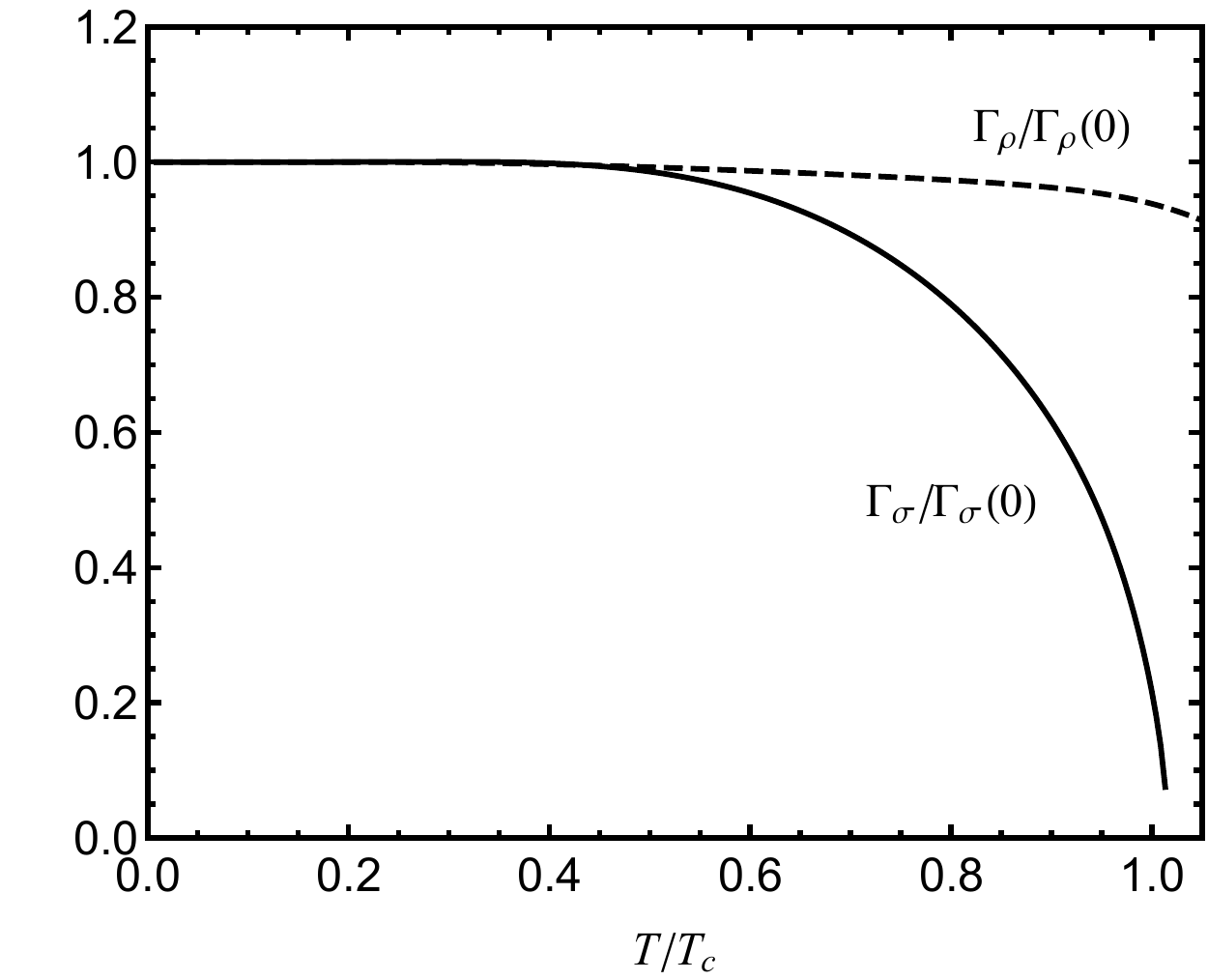}
}
\caption{Left panel: the temperature dependence $M_i(T)$  of mass spectra for mesons ($i=\pi, \sigma, \rho$) and doubled quark mass $m(T)$).
Right panel: scaled temperature dependence of total decay widths $\Gamma_i(T)/\Gamma_i(0)$ for mesons ($i=\sigma, \rho$) calculated within NJL model.
\label{masses}}
\end{figure}

The double quark mass, masses of scalar, pseudoscalar and vector mesons as functions of the temperature at zero chemical potential are presented in Fig.~\ref{masses} (left panel).  
The temperature is normalized to the pseudocritical temperature $T_c = 0.225$ GeV of the chiral crossover transition at zero chemical potential. 
It is clearly seen, that the masses of the $\pi$- and $\rho$-meson stay almost constant when $T<T_c$. Then the pion mass increases and exceeds the mass of two constituent quarks at the Mott temperature $T^\pi_{\rm Mott} = 0.244$ GeV where the dissociation of the pion occurs (also called "soft deconfinement" \cite{Hufner:1996pq}) that is analogous to the Mott effect in solid-state and plasma physics \cite{Zablocki:2010zz}. 

The temperature dependencies for $\Gamma_{\sigma\pi\pi}$ and $\Gamma_{\rho\pi\pi}$, calculated in the framework of the NJL model are shown in the Fig. \ref{masses} (right panel) as black and blue solid lines, accordingly. 
It can be seen that the $\sigma\rightarrow\pi\pi$ decay is forbidden after $T^\sigma_{\rm diss} = 0.228$ GeV, where $\Gamma_{\sigma\pi\pi}$ drops to zero.
A center of mass momentum dependence for $\Gamma_\rho$ and $\Gamma_\sigma$, are introduced by using of Eqs.~(\ref{gsi_prakash}) and (\ref{gro_prakash}). 

\begin{figure}[h]	
\centerline{
\includegraphics[width= 7 cm]{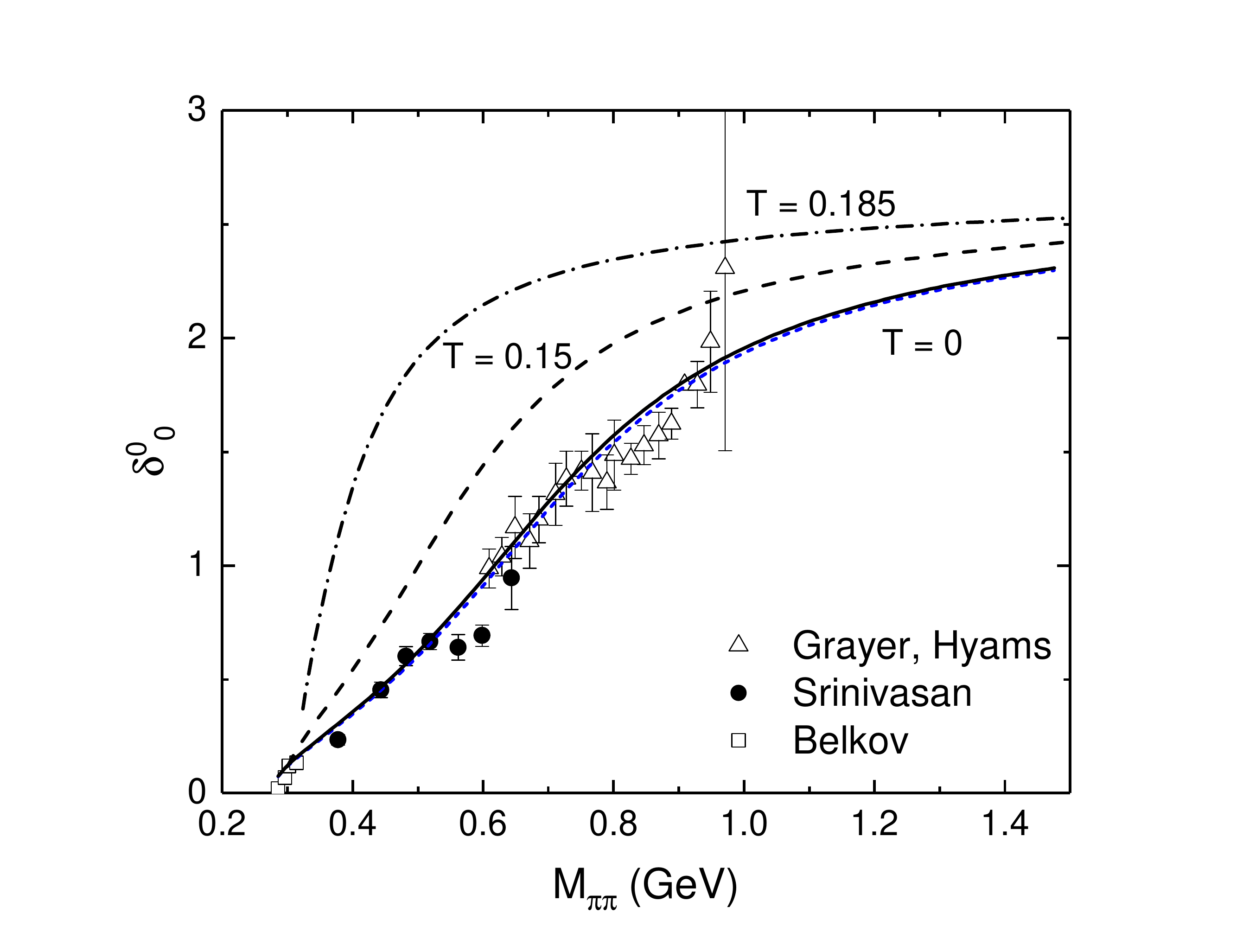}
\includegraphics[width= 7 cm]{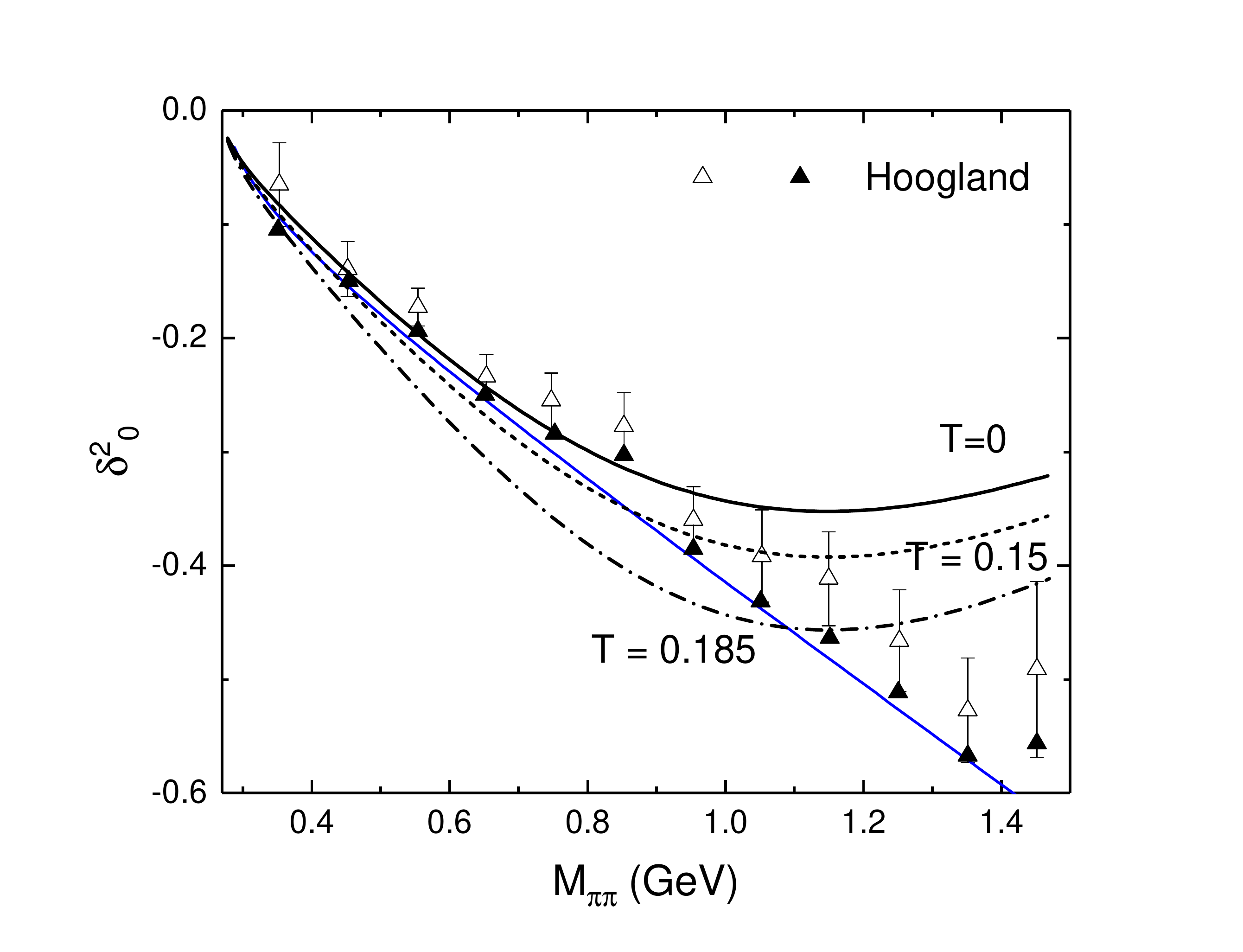}}
\caption{The phase shifts $\delta_0^0$ (left panel) and $\delta_0^2$ (right panel) for different temperatures $T=0$, 0.15, 0.185 GeV. 
The solid blue line in the right panel corresponds to the scattering length approximation $\delta_0^2= -0.12 q/M_\pi$. Experimental data are taken from \cite{Belkov:1979en,Grayer:1974cr,Hyams:1973zf,Srinivasan:1975tj} for $\delta_0^0$ and from \cite{Hoogland:1977kt} for  $\delta_0^2$. }
\label{deltas_T}
\end{figure}

 The phase shift $\delta_0^0$ calculated using 
 Eq.~(\ref{phaseshift00}) for the temperatures   $T=0$, 0.15, 0.185 GeV is shown in left  panel of Fig.~\ref{deltas_T}.   The blue line corresponds to the Eq.~(\ref{phaseshift00}) without T-dependence and the results almost coincide due to the choice of the NJL parameters giving the mass  $M_\sigma \sim 0.8$ GeV. The phase shift $\delta_0^2$ calculated by  Eq.~(\ref{d02Barnes}) for the temperatures  $T=0$, 0.15, 0.185 GeV is shown in the right panel of Fig.~\ref{deltas_T}.  The solid blue line reproduces the scattering length approximation $\delta_0^2= -0.12 q/m_\pi$. The experimental data are taken from \cite{Belkov:1979en,Grayer:1974cr,Hyams:1973zf,Srinivasan:1975tj} for $\delta_0^0$ and \cite{Hoogland:1977kt} for  $\delta_0^2$. 

\begin{figure}[h]	
\centerline{
\includegraphics[width= 7 cm]{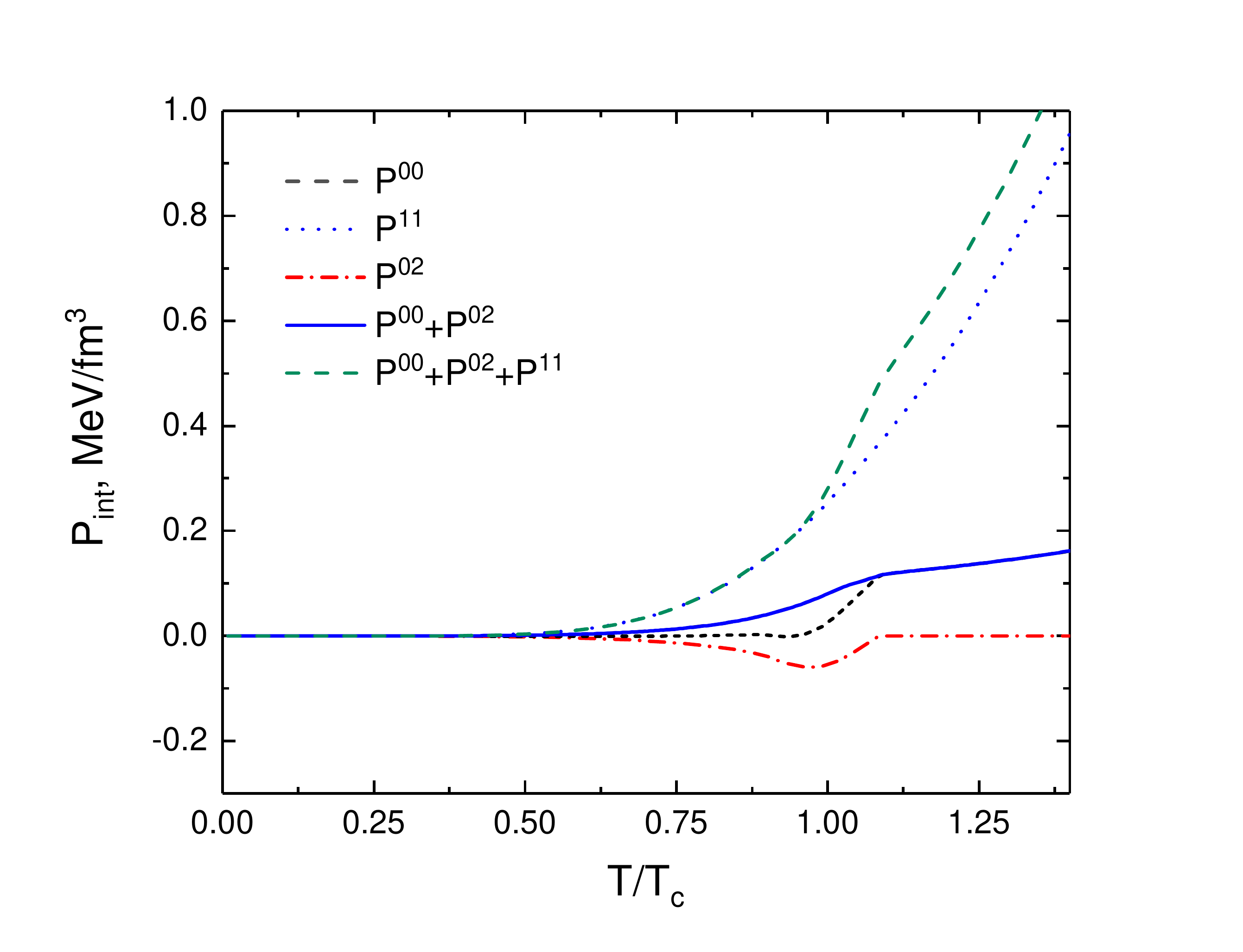}
\includegraphics[width= 7 cm]{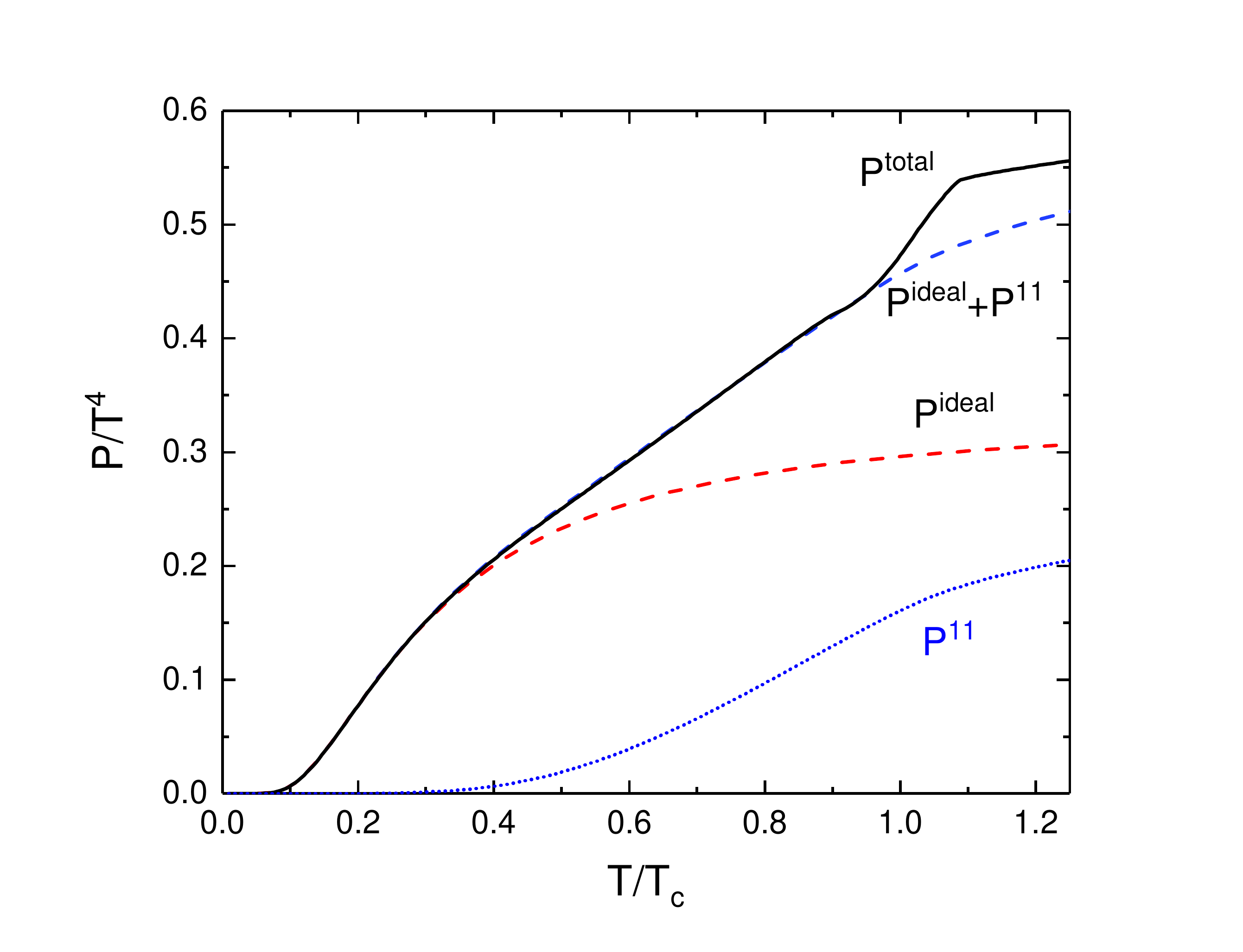}
}
\caption{Left panel: Contributions to the pressure of the pion gas at second order of the virial expansion ($\pi\pi$ scattering) using the Beth-Uhlenbeck equation with medium dependent phase shifts.
The quark Pauli blocking term $P^{02}$ (blue solid line) is modified to take into account that it vanishes when the pion bound state is dissociated. 
Right panel: Pressure of the ideal pion gas (red dashed line), compared to the total pressure with all three interaction channels  
(black solid line) and with just the $\rho$-meson channel (blue dotted line).
The $\rho$-meson contribution is also shown separately by the blue dotted line.
}
\label{pressure}
\end{figure}

The partial contributions to the total pressure for mesons were calculated according to Eq.~(\ref{press_b1}). The repulsive isospin-2 channel phase shift was calculated according to Eq.~(\ref{d02Barnes}). 
The partial contributions to the total pressure are presented in the left panel of the Fig.~\ref{pressure} as functions of the temperature. The contributions from $\delta^0_0$ and $5 \delta^2_0$ compensate each other and the main virial correction to the pressure is given by the $\rho$ meson. 
The pressure of the ideal pion gas (red dashed line) and the interacting pion gas (black solid line) are presented in the right panel of Fig.~\ref{pressure}.

\section{Results and conclusion}
\label{result}

In this paper we discussed that the contribution to the pressure from the S-wave channel is small due to an approximate cancellation of the attractive (I = 0) sigma resonance contribution against the repulsive (I = 2) contribution that is explained by quark Pauli blocking. 
Using the simple Breit-Wigner approximation for the meson phase shifts \cite{Welke:1990za} and the nonrelativistic potential model result \cite{Barnes:1991em} for the phase shift in the repulsive channel (I=2) together with the temperature dependence of the mass spectra from 
the NJL model, we show that this cancellation appears not only in low-energy and low-temperature region as it was discussed before \cite{Venugopalan:1992hy,Broniowski:2015oha}, but also takes place at finite temperature as long as 
$T \lesssim 0.75 T_c$. 

In Table \ref{table1} the relation $a^0_0/(5a^0_2)$ obtained in the frame of NJL model for different temperatures is shown \cite{Quack:1994vc}.   

\begin{table}[h]
\caption{The relation $a^0_0/(5a^0_2)$ in the frame of NJL model.} \label{table1}
\centering
\begin{tabular}{cccccc} 
\hline
 & T=0 & T = 0.14  & T = 0.16 & T = 0.19 & T =  0.2
\\ 
\hline
$a^0_0$ & 0.148 & 0.155  & 0.164  & 0.202  & 0.237\\ 
$a^0_2$ & -0.036 & -0.037  & -0.037  & -0.041 & -0.043 \\
$a^0_0/(5a^0_2)$& -0.826 & -0.85 &-0.88&-0.996 & -1.11\\ \hline
\end{tabular}
\end{table}

Results obtained in the Table~\ref{table1} and  results are shown in the Fig.\ref{pressure} show that the cancellation works also at finite temperature. As can be seen from the Fig.~\ref{pressure}, at high temperatures near the phase transition the difference $P^{00}+5P^{20}$ becomes finite and $\sigma$-meson channel should be taken into account. Similar picture appears in the NJL model, where the scattering length approximation is used \cite{Quack:1994vc}.  It was discussed in Introduction, that singularity in scattering lengths  appears near the $T_c$ due to the cancellation of the input from the $\sigma$-exchange diagram to the total amplitude. Near the critical temperature there appears the interplay between the pion Pauli blocking process and the creation of $\sigma$ meson as a bounding state just before turning back into resonance state \cite{Volkov:1997dx}.

That leads to the fact, that deep inside the HRG phase for $T \ll T_c$ a complete cancellation of the $\sigma$ meson against the repulsive channel appears so that the $\sigma$ meson should not be included to the HRG model \cite{Broniowski:2015oha}. 
In the hadronization region for $T\approx T_c$, however, it is important to take into account the $\sigma$ meson as the degree of freedom \cite{Andronic:2008gu}, since there it is a good resonance and the quark Pauli blocking ceases because of the Mott dissociation of the pion.


\acknowledgments{
D.B. acknowledges discussions with Anton Andronic and Wojciech Broniowski at  Quark Matter 2022 in Cracow.
}








\end{document}